\documentclass[aps,prl,reprint]{revtex4-2}
\usepackage{amsmath,amssymb,amsthm,hyperref,graphicx}
\hypersetup{hidelinks}
\usepackage[capitalise]{cleveref}
\hypersetup{colorlinks=true, linkcolor=blue, citecolor=red, urlcolor=blue}

\begin{document}
	\title{Single-mode spin-wave laser driven by spin-orbit torque}
	\author{J. S. Harms$^{1,3}$}
	\email{joren.harms@uni-konstanz.de}
	\author{H. Y. Yuan$^{1,4}$}
	\author{Rembert A. Duine$^{1,2}$}
	\affiliation{$^{1}$Institute for Theoretical Physics, Utrecht University, 3584CC Utrecht, The Netherlands}
	\affiliation{$^{2}$Department of Applied Physics, Eindhoven University of Technology, P.O. Box 513, 5600 MB Eindhoven, The Netherlands}
	\affiliation{$^{3}$Fachbereich Physik, Universit\"at Konstanz, D-78457 Konstanz, Germany}
	\affiliation{$^{4}$Institute for Advanced Study in Physics, Zhejiang University, 310027 Hangzhou, China}
	\date{\today}
	\begin{abstract}
		A central goal in spintronics and magnonics is the use of spin waves rather than electrons for efficient information processing.
		The key to integrate such spintronic circuits with electronic circuits is the ability to inject, control and detect coherent spin waves with charge currents.
		Here, we propose a tunable setup consisting of a synthetic antiferromagnet in an inhomogeneous magnetic field in which one of the magnetic layers is thin and biased by spin-orbit torque.
		We show that for appriopriate conditions single-mode coherent spin waves are emitted in this set-up. 
		The set-up implements coupling of continuum spin waves with a finite region of negative energy spin waves, such that specific frequencies become self-amplified and thus start lasing.
		We show there exist a large region in parameter space for which the coherent spin wave laser is stabilized by non-linearities and spin-orbit torques.
		Our findings may lead to new ways of injecting coherent spin waves with direct currents.
	\end{abstract}
	\maketitle
	\textit{Introduction.---}
	The emergent fields of magnonics and spintronics propose energy efficient circuits and logic devices using spin waves as information carriers~\cite{chumak_magnon_2015,wolf_spintronics_2001,yuan_quantum_2022}.
	Spin waves have the advantages of low power-consumption and efficient parallel data processing.
	Furthermore, spin waves are promising for unconventional computing applications, due to their inherent non-linear nature~\cite{csaba_perspectives_2017,pirro_advances_2021,chumak_advances_2022}.
	To integrate spintronic circuits with electric circuits, the ability to inject, detect and control magnons using electrical currents is crucial. 
	Although excitation of coherent spin waves with alternating currents (AC) is relatively straightforward, excitation of coherent spin waves using direct currents (DC) is usually more complex and relies on injection of angular momentum with either spin-transfer-torque (STT) or spin-orbit-torque (SOT).
	While excitation of incoherent (thermal) magnons by SOT has been demonstrated~\cite{cornelissen-2016}, efficient excitation of coherent spin waves by DC current is desirable for the technoloqical development of magnonics. 
	
	The first proposals to inject coherent spin waves using DC currents were put forward by Slonczewki~\cite{slonczewski_current-driven_1996} and Berger~\cite{berger_emission_1996,berger_spin-wave_1998}.
	In their proposals, they argued that a spin-polarized current could be used to drive magnetic precession or reorient the magnetization.
	The proposed devices
	are 
	called spin-torque nano oscillators (STNOs) or spin-wave amplification by stimulated emission of radiation (SWASER).
	The injection of angular momentum in these devices gives rise to an angular precession of the free magnetization, which in turn stimulates the emission of spin waves~\cite{hoefer_theory_2005,slavin_spin_2005,slavin_nonlinear_2009}.
	Experimentally, the macrospin precession~\cite{tsoi_excitation_1998}, magnetization reversal~\cite{ozyilmaz_current-induced_2003} and spin wave emission~\cite{rippard_direct-current_2004,demidov_direct_2010,demidov_excitation_2016,madami_direct_2011} in STNOs and SWASERs have been observed.
	Although coherent spin-wave emission using an STNO should be possible, in practice it seems difficult to get single mode coherent spin-wave emission and usually a handfull of modes with a wavelength determined by the size of the STNO are excited
	and
	the frequency of excited spin waves depends on the injected current due non-linearities~\cite{slavin_spin_2005,hoefer_theory_2005}, leading to challenges for its application.
		
	Another type of STNO uses SOT rather than STT to inject spin angular momentum into the magnet.
	The use of SOT rather than STT has the advantage that the spin current that is injected into the magnetic layer is perpendicular to the charge current.
	This allows for the injection of angular momentum in both insulating and conducting magnets, which enables the use of low magnetic damping material such as yttrium iron garnet (YIG).
	Furthermore, SOT can be exerted over a much larger area as compared to STT, leaving room for larger devices.
	On the other hand, SOT-based spin-wave emitters face similar difficulties for injection of coherent spin waves as the STNOs proposed by Slonczewki and Berger~\cite{divinskiy_excitation_2018,demidov_spinorbit-torque_2020,collet_generation_2016}. 
	
	\begin{figure}
		\includegraphics[width=\columnwidth]{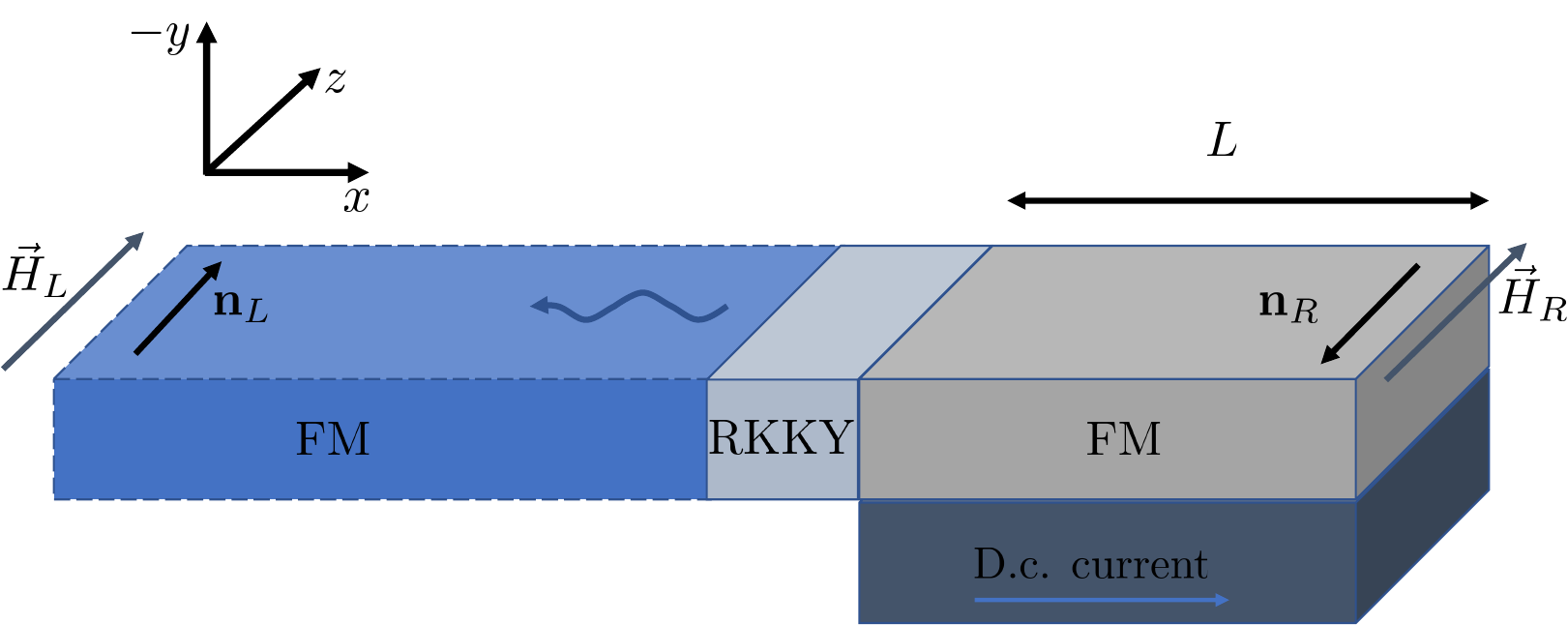}\vspace*{1mm}
		\hspace*{-5mm}
		\includegraphics[width=\columnwidth]{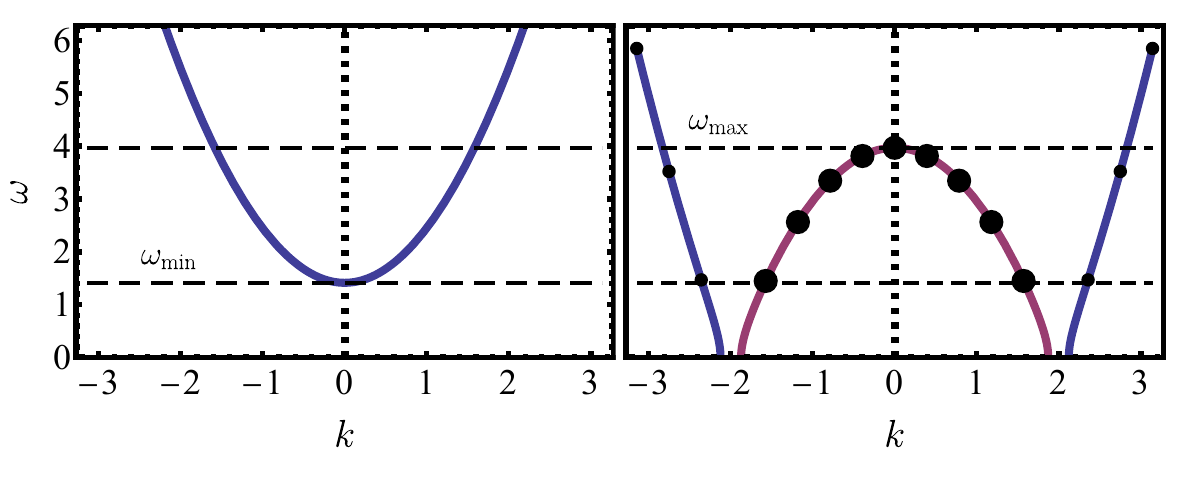}
		\caption{Figure of the setup we propose in this Letter.
		A magnet with magnetization direction along the external field on the left is coupled to a magnet with magnetization direction against the external field on the right.
		The coupling of spin waves in the left magnet to discrete negative energy spin waves in the right magnet facilitates the spin wave laser.
		}
		\label{fig:setup}
	\end{figure}
	
	In this Letter, we propose a setup that excites coherent spin waves via SOT,
	see~\cref{fig:setup}. 
	Here, spin waves are coherently excited by coupling a normal magnet to a confined spatial region with negative energy spin waves.
	This setup has the advantage that it is tunable by the external field and does not experience a non-linear frequency shift depending on the injection of angular momentum.
	
	Besides spontaneous emission of coherent spin waves using STT and SOT~\cite{slonczewski_current-driven_1996,berger_emission_1996,berger_spin-wave_1998,hoefer_theory_2005,slavin_spin_2005,slavin_nonlinear_2009,tsoi_excitation_1998,ozyilmaz_current-induced_2003,rippard_direct-current_2004,demidov_direct_2010,demidov_excitation_2016,madami_direct_2011,slavin_spin_2005,hoefer_theory_2005,divinskiy_excitation_2018,demidov_spinorbit-torque_2020,collet_generation_2016,doornenbal_spin-wave_2019},
	different spin wave laser approaches have been made that amplify spin waves rather than spontaneously emitting them.
	This includes coupling to spin waves to carrier waves~\cite{robinson_spin-wavecarrier-wave_1970,nunes_spin_1982,souto_spin_2001}, quantum amplification~\cite{danilov_effects_1980,danilov_experimental_2002} and the creation of a non-equilibrium distribution using rapid cooling~\cite{breitbach_stimulated_2023}.
	Here, we focus on spontaneous emission of coherent spin waves for which previous setups are discussed above and rely on STTs or SOTs.
	
	Fundamentally, the setup discussed here is quite different from previous STNO or SWASER setups.
	The main difference is that lasing in this setup is induced by the coupling between the magnets rather than driving of the magnet.
	This instability exists due to mode coalescence between positive energy modes in the left magnet and a negative energy mode in the right magnet.
	The negative magnetic energy spin-waves in the right magnet form closed orbits of specific frequencies, which
	become self amplified and hence start lasing.
	The STNOs or SWASERs start lasing due to pumping and thus require driving for the onset of the lasing instability. 
	Contrary, the set-up we propose requires the injection of angular momentum into the right magnet to stabilize the system.

	\textit{Model.---}
	We consider a setup of two exchange coupled ferromagnetic thin insulating films subject to an external magnetic field pointing along the $ z $ direction, see~\cref{fig:setup}. The right film is subject to SOTs keeping the equilibrium magnetization against the direction of the external magnetic field.
	We assume temperatures far below the Curie temperature for which amplitude fluctuations of the magnetization are negligible.
	Accordingly, the dynamics of the magnetization direction $ \mathbf{n}=\mathbf{M}/M_s $ is well described by the Landau-Lifschitz-Gilbert equation (LLG) with spin-orbit torques given by
	\begin{align}\label{eq:LLG}
		\partial_t\mathbf{n}_\nu
		-\alpha\mathbf{n}\times\partial_t\mathbf{n}_\nu
		&=
		-\gamma\mathbf{n}_\nu\times\mathbf{H}_{\mathrm{eff},\nu}
		+I_\nu\mathbf{n}_\nu\times(\hat{z}\times\mathbf{n}_\nu),
	\end{align}
	where $\nu\in\{L,R\}$ denotes the left or right ferromagnet.
	The LLG equation describes damped precession around the effective magnetic field strength $ \mathbf{H}_{\mathrm{eff},\nu}=-\delta E_\nu/(M_s\delta\mathbf{n}) $.
	Here, we consider the magnetic energy functional $ E_\nu[\mathbf{n}] $ in both magnets to be of the form
	\begin{equation}\label{eq:energy-functional}
		\begin{aligned}
			E_\nu=&M_s\int dV \bigg\{	
			\frac{J}{2}(\nabla_i\mathbf{n}_\nu)^2
			-\mu_0H_{\nu}n_{\nu,z}
			\bigg\},
		\end{aligned}
	\end{equation}
	with $ J $ the exchange constant
	and $ H_\nu $ the external magnetic field strength.
	In~\cref{eq:LLG} $ \alpha $ is the dimensionless Gilbert damping constant and $ I_\nu $ characterizes the SOTs --- which are only non-vanishing in the right ferromagnet.
	In principle, anisotropies can relatively straightforwardly be included in the above energy functional, but we do not expect them to change the qualitative physics and omit them here for brevity.
	
	We proceed by defining the canonical coordinate $ \Psi $ to be
	$ \sqrt{2-|\Psi|^2}\Psi=(\hat{x}\mp\mathrm{i}\hat{y})\cdot\mathbf{n} $ and notice that $ n_z=\pm(1-|\Psi|^2) $ -- since the magnetization lives on the sphere.
	In these coordinates the energy functional~\eqref{eq:energy-functional}, up to fourth order in $ \Psi $ and $ \Psi^* $, becomes
	\begin{align}\label{eq:energy-functional-complex-fields}
		E_\nu\simeq M_s\int dV \bigg\{&
		J\left(1-|\Psi_\nu|^2/2\right)(\partial_x\Psi_\nu^*)(\partial_x\Psi_\nu)
		\\\nonumber&
		+\frac{J}{4}\left(\partial_x|\Psi_\nu|^2\right)^2
		\mp\mu_0H_{\nu}\left(1-|\Psi_\nu|^2\right)
		\bigg\}.
	\end{align}
	To incorporate the effect of dissipation, that is, Gilbert damping and SOT, in these canonical coordinates we introduce the Rayleigh dissipation functional
	\begin{align}
		\mathcal{W}_\nu
		=&
		\frac{M_s}{\gamma}\int\mathrm{d}V
		\Big\{
		\frac{\alpha}{2}(\partial_t\mathbf{n}_\nu)^2-J_\nu\hat{z}\cdot(\mathbf{n}_\nu\times\partial_t\mathbf{n}_\nu)
		\Big\}
		\\\nonumber
		=&\frac{M_s}{\gamma}\int\mathrm{d}V
		\big\{
		\alpha\left[1-|\Psi_\nu|^2/2\right](\partial_t\Psi_\nu^*)(\partial_t\Psi_\nu)
		\\\nonumber
		+&\frac{\alpha}{4}(\partial_t|\Psi_\nu|^2)^2
		\pm \mathrm{i}I_\nu\left[1-|\Psi_\nu|^2/2\right]\left(\Psi_\nu\partial_t\Psi_\nu^*-\Psi_\nu^*\partial_t\Psi_\nu\right)
		\big\}.
	\end{align}
	The Rayleigh dissipation functional is also expanded up to fourth order in the fields $ \Psi_\nu $ and $ \Psi_\nu^* $.
	The Euler-Lagrange equations with Rayleigh dissipation yield the two equations of motion describing the non-linear magnetization dynamics
	\begin{equation}
		\begin{aligned}\label{eq:non-linear-equation-of-motion}
		{\mathrm{i}\partial_t\Psi_\nu}
		\simeq
		-&\Lambda^2(1-|\Psi_\nu|^2/2)\partial_x^2\Psi_\nu
		-\Lambda^2\big(\partial_x^2|\Psi_\nu|^2\big)\Psi_\nu/2
		\\
		-&\Lambda^2(\partial_x\Psi_\nu^*)(\partial_x\Psi_\nu)\Psi_\nu/2
		\pm h\Psi_\nu
		\\
		+&\alpha\left[1-|\Psi_\nu|^2/2\right](\partial_t\Psi_\nu)
		+\alpha(\partial_t|\Psi_\nu|^2)\Psi_\nu/2
		\\
		\pm&\mathrm{i}I_\nu\left[1-|\Psi_\nu|^2/2\right]\Psi_\nu,
	\end{aligned}
	\end{equation}
	where the second equation of motion is given by the complex conjugate of the above.
	In the above we defined the dimensionless time $t\rightarrow t/\gamma\mu_0M_s $, the exchange length $\Lambda=\sqrt{J/\gamma\mu_0M_s}$, the dimesionless SOT $ I_\nu/\gamma\mu_0M_s\rightarrow I_\nu $ and the dimensionless magnetic field $h_\nu=H_\nu/M_s$.
	
	We find the linear spin-wave excitations in the left and right magnet using a plane wave ansatz and linearizing the equation of motion. 
	This gives the following dispersion relation for spin waves
	\begin{subequations}\label{eq:linearised-dispersion}
		\begin{align}\label{eq:linearised-dispersion-left-magnet}
			\omega_{L}&=\Lambda^2(k^{L})^2+ h_{L}-\mathrm i \alpha \mathrm {Re}\,\omega_{L},\\
			\omega_{R}&=\Lambda^2(k^R)^2- h_{R}-\mathrm{i}(I_R+\alpha\mathrm {Re}\,\omega_R),\label{eq:linearised-dispersion-right-magnet}
		\end{align}
	\end{subequations}
	where $\omega_{L/R}$ gives the dispersion in the left and right magnet respectively with $k^L\in\mathbb R$ and $ k^R=\pi m/L $ with $ m\in\mathbb{N} $ --- since the right ferromagnet has finite size.
	Dynamical stability of the right ferromagnet, at the linear level, in the absence of coupling between the left and right magnet, requires $\mathrm{Im}\,\omega_R <0$.
	We thus find that the SOT should satisfy $I_R>\alpha h_R$ for the right magnet to be dynamically stable.
	This reversal of the magnetization against the external effective field with electric currents has been demonstrated experimentally in magnetic nanopillars~\cite{kent-2003}.
	From this point onward we ignore Gilbert damping in the left magnet
	and, for future purposes, note that the wavenumber of the left moving mode in the left magnet at the precessional frequency $ \omega $ is given by $ \Lambda k^L_l=-\sqrt{\omega-h_L} $.
	
	Next, we determine the boundary conditions between the continuum of magnons in the left ferromagnet and the discrete states in the energetically unstable right ferromagnet.
	The interaction between the magnets is given by the Ruderman–Kittel–Kasuya–Yosida (RKKY) interaction energy
	\begin{align}
		&E_\mathrm{int}
		=
		-J_c\mathbf{n}_L\cdot\mathbf{n}_R
		\rvert_{x=0}
		=
		J_c\Big[
		(1-|\Psi_L|^2)(1-|\Psi_R|^2)
		\\\nonumber
		&-\sqrt{(1-|\Psi_L|^2/2)(1-|\Psi_R|^2/2)}(\Psi_L\Psi_R+\Psi_L^*\Psi_R^*)
		\Big]\rvert_{x=0},
	\end{align}
	with $J_c$ the RKKY interaction strength.
	From here we find the boundary conditions by employing the variational derivative at the boundaries.
	This yields the boundary condition at the interface between left magnet and the non-magnetic spacer
		\begin{align}\label{eq:non-linear-boundary-conditions}
			&J(1-|\Psi_L|^2/2)\partial_x\Psi_L+J\Psi_L\partial_x|\Psi_L|^2/2
			=\\\nonumber
			&J_c
			\Big[
			\sqrt{(1-|\Psi_L|^2/2)(1-|\Psi_R|^2/2)}\Psi_R^*
			+(1-|\Psi_R|^2)\Psi_L
			\\&\nonumber
			-\sqrt{{(2-|\Psi_R|^2)}/{(2-|\Psi_L|^2})}
			(\Psi_L\Psi_R+\Psi_L^*\Psi_R^*){\Psi_L}/{4}
			\Big].
		\end{align}
	The boundary condition at the interface between the right magnet and the non-magnetic spacer is similar with $\Psi_L\leftrightarrow\Psi_R$ and $\partial_x\rightarrow-\partial_x$. 
	
	We proceed by treating the spin wave fluctuations in the left ferromagnet to be much smaller in amplitude than the amplitude of the standing wave in the right ferromagnet.
	This assumption follows from energy conservation at the weakly coupled interface and by noting that we only allow for outgoing modes in the left magnet.
	As a result we treat the continuum in the left ferromagnet using linear spin-wave theory, while treating the right ferromagnet non-linearly.
	We furthermore assume $ -\Lambda k_l^L\gg\Lambda_c $, which corresponds to weak coupling. The boundary condition in~\cref{eq:non-linear-boundary-conditions}, for weak coupling strengths, becomes
		$
		J\partial_x\Psi_L\rvert_{x=0}
		\simeq
		J_c
		\sqrt{1-{|\Psi_R|^2}/{2}}\Psi_R^*
		\rvert_{x=0}.
		$
	This leaves us with
	\begin{align}\label{eq:left-boundary-condition-result}
		&\mathrm{i}\Lambda k_l^L\Psi_L\rvert_{x=0}
		\simeq
		\Lambda_c
		\sqrt{1-{|\Psi_R|^2}/{2}}\Psi_R^*
		\rvert_{x=0},
	\end{align}
	with $\Lambda_c=J_c/\Lambda\gamma\mu_0M_s$.
	
	Rather than incorporating the RKKY interaction at the right interface as a boundary condition, we include this interaction as a boundary term in the equation of motion of the right ferromagnet after which we substitute~\cref{eq:left-boundary-condition-result}.
	The contribution of the RKKY interaction to the equation of motion of $\Psi_R$ then gives
	\begin{align}
		{\delta E_\mathrm{int}}/{\delta\Psi^*}
		=&
		\big[
		\Lambda_c\Lambda\Psi_R
		-\mathrm{i}({\Lambda_c^2}/{ k_l^L}) (1-|\Psi_R|^2/2)\Psi_R
		\big]\delta(x),
	\end{align}
	which should be added to the equation of motion in~\cref{eq:non-linear-equation-of-motion}.
	The first term in the above describes an energetic boundary contribution, while the second term gives the flow of energy from the right ferromagnetic to the left ferromagnet. 
	
	\textit{Non-linear analyses of the spin-wave lasing mode.---}
	Because the RKKY coupling between the left and right magnet is small compared to their exchange coupling, it is natural to consider a mode expansion which satisfies the exchange boundary conditions, $\partial_x\Psi\rvert_{x\in \{0,L\}}=0$, at the boundaries in the right magnet
	\begin{align}\label{eq:ansatz-psi0}
		\Psi_R=\sum_mA_m(t)e^{\mathrm{i}\phi_m(t)}e^{-\mathrm{i}\omega_mt}\sqrt{\frac{2-\delta_{m,0}}{L}}\cos[k_mx],
	\end{align}
	with $ k_m=\pi m/L $ for $ m\in\mathbb{N} $ and $\omega_m=\Lambda^2k_m^2-h_R$ the precessional frequency without damping of the right magnet.
	We note that the energetic boundary contribution of the RKKY interaction could in principle be included using the Sturm-Liouville expansion.
	We however disregard this.
	Since we consider $\alpha$, $I_R$ and $\Lambda_c^2$ to be small, there is no need to include this correction up to first order in these parameters.
	In the above expansion, we explicitly expect the timescale on which $ A_n $ and $ \phi_n $ change to be much larger than the timescale set by $ \omega_n^{-1} $.
	
	We are ultimately interested in the possibility of limit cycles at finite amplitude because they will correspond, as we shall see, to lasing or coherent emission of spin waves. 
	We start with the approximation in which only one non-uniform mode is present.
	For a non-uniform mode, $n\neq0$, the equations of motion in~\cref{eq:non-linear-equation-of-motion} become
	\begin{subequations}
		\begin{align}\label{eq:amplitude-equation-single-mode-a}
			\partial_t\phi_{n}
			\simeq&
			-\alpha (\partial_TA_{n})(1+3A_{n}^2/4L)
			+2\Lambda_c\Lambda/L,
			\\\label{eq:amplitude-equation-single-mode-b}
			\partial_tA_{n}
			\simeq&
			-(2\Lambda_c^2/k_l^LL)(1-A_{n}^2/L)A_{n}
			\\\nonumber&
			-(\alpha\omega_{n}-\partial_t\phi_{n}+I_R)(1-3A_{n}^2/4L)A_{n}.
		\end{align}
	\end{subequations}
	The first equation gives corrections to the precessional frequency due to amplitude fluctuations and the RKKY interaction.
	The second equation describes the dissipative dynamics of the amplitude of the $n$-th spin-wave mode.
	As we have seen before, the linear equation of motion in~\cref{eq:linearised-dispersion-right-magnet} predicts that $I_R>\alpha h_R$ for the spin waves in the right ferromagnetic to be stable to start with.
	Furthermore,~\cref{eq:amplitude-equation-single-mode-b} predicts the onset of a linear instability for coupling strengths $\Lambda_c$ exceeding
	$
	-2\Lambda_c^2/k_{l,n}^LL>I_R+\alpha\omega_{n},
	$
	where we used the shorthand notation $k_{l,n}^L\equiv k^L_l(\omega_n)=-\sqrt{\omega_n-h_L}$.
	Hence, the closed orbits of negative energy spin waves become self amplified for sufficiently strong coupling strengths.
	This instability is due to mode coalescence between spin waves in the left ferromagnet and the negative energy standing waves in the right magnet~\cite{coutant_black_2010}.	
	The mechanism behind this is similar to the formation of an exceptional point~\cite{coutant_dynamical_2016}.
	Including the non-linear contributions in~\cref{eq:amplitude-equation-single-mode-b}, we find the amplitude of the self amplified mode to be stabilized by the non-linearities and to be given by  
	\begin{align}\label{eq:amplitude-lasing-mode}
		\frac{A_{n}^2}{L}
		\simeq
		\frac{x_{n}-1}
		{x_{n}-3/4},
	\end{align}
	with
	$
	x_{n}=-{2\Lambda_c^2}/{(\alpha\omega_{n}+I_R)k^L_{l,n}L}>1.
	$
	
	We may express the emitted spin current in terms of the above spin wave amplitude.
	This can be done since the amplitude of emitted spin waves is related to the amplitude of the spin wave mode in the right magnetic via~\cref{eq:left-boundary-condition-result}.
	We further define the dimensionless spin-current in the left ferromagnet by $\mathrm{i}J_\mathrm{spin}=\Psi_L\partial_x\Psi^*_L-\Psi^*_L\partial_x\Psi_L$.
	This yields that the coherent spin current carried by spin waves of frequency $\omega_n$ emitted by this lasing setup is given by
	\begin{align}
		J_\mathrm{spin}
		=|A_L|^2\Lambda k^L_{l,n},
	\end{align}
	with $|A_L|^2=2(\Lambda_c/\Lambda k_{l,n}^L)^2(1-{A_{n}^2}/{L})A_{n}^2/L$.
	
	In the proceeding section we consider the DC current interval ---which generates the SOT--- for which this single-mode laser remains stable.
	
	\textit{Current interval for a stable single-mode laser.---}
	From this point onwards we consider the $n$-th mode to be the linearly most unstable mode and the interaction between this lasing mode and the other modes close to the resonant condition $\omega_{n'}=\omega_{n_1}-\omega^*_{n_2}+\omega_{n_3}$, with $n_1$, $n_2$ and $n_3$ arbitrary for the moment.
	Since by assumption the $n$-th mode is the only mode with non-vanishing amplitude, we consider $n_2=n$ and either $n_1=n$ and $n_3=n'$ or $n_1=n'$ and $n_3=n$.
	We therefore consider only modes that appear twice in~\cref{eq:non-linear-equation-of-motion}.
	To recap, we are interested in the stability of the situation in which the $n$-th mode is lasing and the other modes remain stable, in the sense that their amplitude remains vanishing.
	The equation of motion for the amplitude of these other modes follows from~\cref{eq:non-linear-equation-of-motion,eq:left-boundary-condition-result} as
	\begin{align}\label{eq:amplitude-constraint-1}
		&\frac{\partial_tA_{n'}}{A_{n'}}
		\simeq
		-\alpha\omega_{n'}
		- I_R
		+\frac{(2-\delta_{n'})\Lambda_c^2}{ k_{l,n'}^L L}
		\\\nonumber&
		+\left[
		\alpha
		\omega_{n}
		+ I_R
		+\frac{\Lambda^2 k_{n}^2}{2}
		-\frac{2(2-\delta_{n'})\Lambda_c^2}{k_{l,n'}^L L}
		\right]\frac{A_{n}^2}{L}<0,
	\end{align}
	where the last equality is the stability requirement.
	Hence, the one mode laser is stable if the amplitude in~\cref{eq:amplitude-lasing-mode} satisfies the above constraint for arbitrary $n'$.
	We expect the uniform mode or the first non-uniform be the most susceptible to the instability.
	Therefore we consider $ n'\rightarrow 0$ in the following. 
	With use of~\cref{eq:amplitude-lasing-mode}, the constraint in~\cref{eq:amplitude-constraint-1} is written as a quadratic equation in $I_R-\alpha_R+\alpha\Lambda^2k_n^2$.
	Namely,
	$
	(I_R-\alpha h+\alpha \Lambda^2 k_{n}^2)^2
	+(I_R-\alpha h+\alpha \Lambda^2 k_{n}^2)
	[
	5\alpha \Lambda^2 k_{n}^2
	+{5\Lambda_c^2}/{k_{l,0}^L L}
	]
	+
	({4\Lambda_c^2}/{k^L_{l,n}L})
	[
	{2\Lambda_c^2}/{k_{l,0}^L L}
	+3\alpha \Lambda^2 k_{n}^2
	]
	\gtrsim0.
	$
	From the fact that this equation is quadratic we find that all currents within the linearly unstable range $-{2\Lambda_c^2}/{k^L_{l,n}L}>I_R+\alpha\omega_{n}>0$ are in principle allowed if the discriminant is negative.
	By assuming $1/k_{l,n'}^L L$ to be small compared to $1/k_{l,n}^L L$ we find the critical coupling strength to be approximately
	\begin{equation}
		\begin{aligned}
			\Lambda_{c,\mathrm{critical}}^2
			\gtrsim
			3\alpha\Lambda^2k_{n}^2|k^L_{l,0}|L/2.
		\end{aligned}
	\end{equation}
	\begin{figure}
		\centering
		\includegraphics[width=.9\columnwidth]{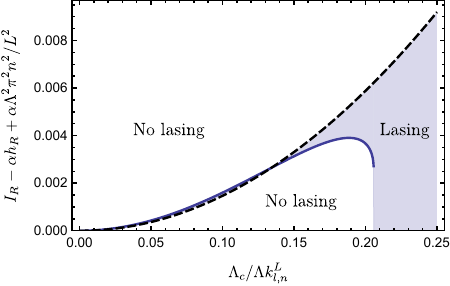}
		\caption{Figure of the current interval presented in~\cref{eq:maximal-current}, for $h_R-h_L=1/2$, $L/\Lambda=8$ and $\alpha=10^{-2}$.
			For these parameters, the lasing mode occurs at $n=1$.
			The solid curve describes the lower bound, while the dashed curve describes the upper bound.
			For sufficiently large interaction strengths between the magnets we find a large DC current interval in which this setup is a stable single-mode laser.
			We furthermore notice a sudden drop in the required current $I_R$ for coupling strengths $\Lambda_c>\Lambda_{c,\mathrm{critical}}$.
			For small interaction strengths, the upper and lower bound swap and hence there is no possibility of a stable single-mode laser in this regime.
			}
		\label{fig:current-interval}
	\end{figure}\noindent
	If $\Lambda_c$, on the other hand, is smaller than this critical value, the current interval in which there exists a stable one mode laser becomes
	\begin{align}\label{eq:maximal-current}
		&-{2\Lambda_c^2}/{k^L_{l,n}L}\nonumber
		\geq
		I_R-\alpha h+\alpha \Lambda^2 k_{n}^2
		\gtrsim
		{5}\left(
		\alpha \Lambda^2 k_{n}^2
		+\Lambda_c^2/{k_{l,0}^L L}
		\right)
		\\&
		\times
		\frac{1}{2}
		\left[
		\sqrt{
			1	
			-\frac{48\Lambda_c^2}{25}
			\frac{k_{l,0}^L }{k^L_{l,n}}
			\frac
			{\alpha \Lambda^2 k_{n}^2k_{l,0}^L L+{2\Lambda_c^2/3}}
			{(\alpha \Lambda^2 k_{n}^2k_{l,0}^L L+\Lambda_c^2)^2}
		}
		-1
		\right].
	\end{align}
	The upper bound in the injected current comes from the lasing condition $ -{2\Lambda_c^2}/{k^L_{l,n}L}\geq I_R-\alpha h+\alpha \Lambda^2 k_{n}^2$ which makes the $A_{n}=0$ an unstable fixed point in~\cref{eq:amplitude-equation-single-mode-b}.
	In~\cref{fig:current-interval}, we plot the electrinic DC current interval in~\cref{eq:maximal-current} as a function of the coupling strength $\Lambda_c$.
	We like to stress that the lower boundary in the current $I_R$ for $\Lambda_c>\Lambda_{c,\mathrm{critical}}$ becomes $I_R>\alpha(h_R-\Lambda^2k_n^2)$, i.e. the required current for negative energy spin waves ---with frequency $\omega_n$--- to be dynamically stable.
	We find this lowest current to be lower than the current needed to stabilize the linear spin waves in~\cref{eq:linearised-dispersion-left-magnet}.
	
	Furthermore,~\cref{eq:amplitude-lasing-mode,eq:amplitude-constraint-1,eq:maximal-current} give a lower bound in the interfacial coupling strength below which no current interval that stabilizes the one mode laser exists.
	We find this lower bound in the coupling strength to be
	\begin{equation}
		{2\Lambda_{c,\mathrm{lower~bound}}^2}\gtrsim\alpha\Lambda^2k_{n}^2|k^L_{l,n}|L.
	\end{equation}
	Let us finish by checking the self-consistency of the assumption $-\Lambda k_{l,n}^L\gg\Lambda_c$, which allows us to treat the spin waves in the left domain linearly.
	This assumption implies
	$1\gg\Lambda_c^2/(\Lambda k_{l,n}^L)^2\gtrsim\alpha k_{n}^2L/2|k^L_{l,n}|$, which is equivalent to
	$
	{(2L/\Lambda)\sqrt{h_R-h_L-n^2\pi^2\Lambda^2/L^2}}
	\gg {\alpha n^2\pi^2}{}.
	$
	Hence, a single-mode laser is stable if only a few --- and most likely only two or three --- energetically unstable modes are present.

	\textit{Conclusion, discussion and outlook.---}
	In this Letter, we proposed a way to coherently inject spin-waves of a specific frequency, thereby constituting a spin wave laser driven by a DC current.
	This realization depended crucially on the coupling of magnonic excitations to negative energy magnon excitations in a confined region. 
	Via the coupling of magnons to these negative energy magnons, the system can form closed orbits, thereby dynamically destabilizing the system.
	The formation of the dynamical unstable modes is similar to that of the mode coalescence forming an exceptional point.
	In the literature of analogue gravity such a set-up is a realization of a black-hole laser~\cite{corley_black_1999}.
	Since finite many modes start lasing, i.e., exponentially growing, non-linearies quickly become important.
	Hence, we investigated the non-linear regime for which the single-mode laser is stable.
	More specifically, we considered the case in which one mode dominates and analyzed its stability towards instabilities in other modes.
	We found there to be a region in parameter space for which this single-mode spin wave laser is stable and emits coherent spin waves.
	We consider typical experimental values in Yttrium Iron Garnet (YIG) for the saturation magnetization $\mu_0H\sim\mu_0M_s\sim0.25~T$, the gyromagnetic ratio $ \gamma/2\pi\sim30\mathrm{GHz\,T^{-1}} $ and the Gilbert damping $\alpha = 10^{-4}$.
	The SOT is related to the electric current density $J_c$ via
	$
	I_R\sim
	\gamma\hbar\eta\theta_\mathrm{SH}J_c/2eM_sd
	$~\cite{manchon2019current}, with $e$ the electron charge, $\eta~\sim\mathcal{O}(1)$ the interfacial efficiency, $d\sim2\,\mathrm{nm}$ the thickness of the ferromagnet and $\theta_{SH}=0.1$ the spin-Hall angle for Platinum \cite{wang2014scaling}.
	The current necessary to keep this setup in the lasing regime is of the order $I_R/\gamma\mu_0M_s\sim\alpha H_R/M_s\sim\mathcal{O}(\alpha)$.
	Hence, we estimate the current density needed to stabilize the laser to be of the order $J_c\sim10^{8}\,\mathrm{A/m^2}$.
	Furthermore, the exchange length for YIG is approximately $\Lambda\sim9~\mathrm{nm}$ and the dimensionless RKKY coupling is tunable and of the order $\Lambda_c\sim\mathcal{O}(1)$~\cite{wang2023spin}.
	The results in this Letter have been found assuming weak coupling between the magnets, i.e. $ -\Lambda k_l^L\gg\Lambda_c $, which allows us to treat spin waves in the left magnet linearly. This approximation becomes less exact when the resonant frequencies in the right magnet are close to the Kittel frequency of the left magnet or when exploring stronger coupling strengths.
	Developing a theory beyond this limit is in principle of interest, since the onset of lasing is easier to achieve for large couplings.
	In future work, one could develop a theory that treats the non-linear magnetization dynamics of the left magnet.
	Furthermore, it could useful to consider this setup with two interfaces in the quantum regime.
	This is due to the fact that this setup is likely to produce entangled pairs of magnons.
	Another direction would be to consider the coupling between positive and standing wave negative energy magnons in antiferromagnets to create a single-mode laser in this class of materials too.
	\section*{Acknowledgements}
	R. A. Duine is member of the D-ITP consortium, a program of the Netherlands Organisation for Scientific Research (NWO) that is funded by the Dutch Ministry of Education, Culture and Science (OCW). R.A.D. acknowledges the funding from the European Research Council (ERC) under the European Union's Horizon 2020 research and innovation programme (Grant No. 725509).
	This work is part of the Fluid Spintronics research programme with project
	number 182.069, which is financed by the Dutch
	Research Council (NWO).
	H.Y. Yuan is supported by the National Key R$\&$D Program of China (2022YFA1402700) and Marie Sk{\l }odowska-Curie Grant Agreement SPINCAT (101018193).
	\bibliography{BH-laser}
\end{document}